# Shared rewarding overcomes defection traps in generalized volunteer's dilemmas

Xiaojie Chen[a,1], Thilo Gross[b], and Ulf Dieckmann[a]

**Author affiliations**

[a]Evolution and Ecology Program, International Institute for Applied Systems Analysis (IIASA), Schlossplatz 1, A-2361 Laxenburg, Austria

[b]Department of Engineering Mathematics, Merchant Venturers Building, University of Bristol, Bristol BS8 1UB, United Kingdom

[1]Corresponding author: email chenx@iiasa.ac.at, phone +43 2236 807 321

**Abstract**

For societies to produce or safeguard public goods, costly voluntary contributions are often required. From the perspective of each individual, however, it is advantageous not to volunteer such contributions, in the hope that other individuals will carry the associated costs. This conflict can be modeled as a volunteer's dilemma. To encourage rational individuals to make voluntary contributions, a government or other social organizations can offer rewards, to be shared among the volunteers. Here we apply such shared rewarding to the generalized $N$-person volunteer's dilemma, in which a threshold number of volunteers is required for producing the public good. By means of theoretical and numerical analyses, we show that without shared rewarding only two evolutionary outcomes are possible: full defection or coexistence of volunteers and non-volunteers. We show that already small rewards destabilize full defection, stabilizing small fractions of volunteers instead. Furthermore, at these intermediate reward levels, we find a hysteresis effect such that increasing or decreasing group sizes can trigger different social outcomes. In particular, when group size is increased, the fraction of volunteers first increases gradually before jumping up abruptly; when group size is then decreased again, the fraction of volunteers not only remains high, but even continues to increase. As the shared reward is increased beyond a critical level, the bistablitity underlying this hysteresis effect vanishes altogether, and only a single social outcome remains, corresponding to the stable coexistence of volunteers and non-volunteers. We find that this critical level of shared rewarding is relatively small compared to the total cost of contributing to the public good. These results show that the introduction of shared rewarding is remarkably effective in overcoming defection traps in the generalized volunteer's dilemma.

**Keywords**

# 1. Introduction

Understanding the emergence and stability of cooperation among rational individuals is a central challenge in evolutionary biology, as well as in the social sciences. Evolutionary game theory provides a common mathematical framework for interpreting the evolution of cooperation. In particular, the prisoner's dilemma game and its $N$-person variant are the most commonly employed games for studying this challenge (Axelrod and Hamilton, 1981; Maynard Smith, 1982; Hauert et al., 2002; Doebeli and Hauert, 2005; Nowak, 2006; Hauert et al., 2007; Szabó and Fáth, 2007; Santos et al., 2008; Traulsen et al., 2009; Sigmund et al., 2010; Perc et al., 2013). However, there are some social dilemmas concerning altruistic behavior for which different games offer more appropriate models. In particular, the volunteer's dilemma game has been proposed as another important paradigm (Diekmann, 1985).

The volunteer's dilemma is defined as follows (Diekmann, 1985): members of a group can volunteer to pay a small cost (Volunteer, a cooperative strategy) or avoid to pay the cost (Ignore, a defective strategy) towards the production or maintenance of a public good. If at least one individual chooses to volunteer, the public good is produced and benefits all individuals in the group, irrespective of their contributions. In contrast, if nobody volunteers, the public good is not produced, and all group members pay a cost that is higher than that of volunteering. Clearly, a volunteer benefits from his or her action if nobody else volunteers, whereas such a voluntary investment is wasted if another group member volunteers as well (Archetti, 2009b). This volunteer's dilemma game has been extended to the more general case in which more than one volunteer is required for producing the public good; the resultant game is also known as a threshold public good game (Myatt and Wallace, 2008; Archetti, 2009a; Boza and Számadó, 2010).

The volunteer's dilemma and its generalization can be applied to many situations studied in the social sciences (Diekmann, 1985; Diekmann, 1993), such as volunteering work in charitable organizations, cleaning shared accommodation, or getting up at night to placate a crying baby (Bilodeau and Slivinski, 1996; Otsubo and Rapoport, 2008). Moreover, the volunteer's dilemma is relevant also in the context of evolutionary biology (Goeree and Holt, 2005; Archetti, 2009a,b; Archetti and Scheuring, 2011; Archetti, 2011): a typical example from the biological context is a population of social animals using alarm calls to warn others of predators. In such a situation, the individual raising the alarm promotes the collective security of its population from predation, but individually incurs non-negligible costs, because raising an alarm often increases the risk of being targeted by a predator.

Previous studies have examined different factors affecting cooperation in volunteer's dilemmas, such as group size (Franzen, 1995; Weesie and Franzen, 1998; Archetti, 2009b; Brännström et al., 2011), individual vigilance (Archetti, 2011), and nonlinear benefits (Do et al., 2009; Archetti and Scheuring, 2011). Specifically, it was found that the fraction of volunteers decreases with group size, so that larger groups tend to under-produce the public good (Franzen, 1995; Weesie and Franzen, 1998; Archetti, 2009b; Brännström et al., 2011). Above a certain group size, reducing an individual's vigilance can induce other players to volunteer more often (Archetti, 2011). Moreover, incorporating nonlinear returns on investments allows cooperation to be sustained, so that cooperators and free-riders are expected to coexist in a stable mixed equilibrium (Archetti and Scheuring, 2011).

In this study, we incorporate a rewarding mechanism into the volunteer's dilemma. Our motivation for this extension stems from the observation that in many real-world situations voluntary contributions are maintained by a reward system providing incentives for volunteering. For example, companies or enterprises reward groups of employees for good performance in teamwork, and volunteers at events such as Olympic Games receive favorable public recognition. Furthermore, governments or other social organizations involved with public security enact regulations that reward and protect citizens who voluntarily strengthen the fight against crime (Marin and Harder, 1994). These volunteers can thus become role models for other people, which further stimulates voluntary behavior.

In this study, we assume that volunteers in a group receive a certain reward from an external pool of resources that is shared among them. We study how the introduction of such shared rewarding influences the equilibrium fraction of volunteers in large well-mixed populations, and what level of rewarding is needed to overcome defection traps in these systems. We find that the introduction of shared rewarding has two interesting consequences. First, shared rewarding leads to a hysteresis effect under which the highest level of volunteering is reached by first increasing and then decreasing group size. Second, even small total rewards suffice to stabilize the coexistence of volunteers and non-volunteers, and are thus surprisingly efficient in robustly overcoming defection traps.

## 2. Methods

### 2.1. Volunteer's dilemma

We consider the generalized $N$-person volunteer's dilemma, describing the interaction of individuals in groups of $N$ ($N \geq 2$) players. In each round of the game, an interaction group is assembled by randomly drawing $N$ individuals from a large (infinite) well-mixed population (Hauert et al., 2002; Hauert et al., 2006). Each player can choose between the strategies Volunteer or Ignore. The public good is produced if at least $k$ ($1 \leq k < N$) individuals choose to volunteer. The cost of volunteering is $c$, relative to a baseline payoff of 1; volunteers incur this cost irrespective of whether or not the public good is produced. A failure of producing the public good imposes a cost $a > c$ on each player in the group. As a result, the payoffs of the strategies Volunteer and Ignore are given by

$$P_V(M) = \begin{cases} 1 - c & \text{if } M \geq k, \\ 1 - a - c & \text{if } M < k, \end{cases} \qquad (1a)$$

and

$$P_I(M) = \begin{cases} 1 & \text{if } M \geq k, \\ 1-a & \text{if } M < k, \end{cases} \quad (1b)$$

respectively, where $M$ is the number of volunteers in the group. For $c > 0$, $P_I(M)$ thus always exceeds $P_V(M)$.

It is worth pointing out that payoffs in the volunteer's dilemma can alternatively be described by assuming a cost resulting from the public good's absence (as above) or a benefit resulting from its presence (Archetti, 2011). The evolutionary dynamics reported below are unaffected by this alternative parameterization.

## 2.2. Shared rewarding

We extend the $N$-person volunteer's dilemma specified above by introducing shared rewarding. For this we assume that the volunteers in an interaction group in each round share a total reward $\delta$ ($0 < \delta < Nc$). Whether $\delta$ is provided from an external pool of resources or through compulsory contributions made by all players has no bearing on the evolutionary dynamics reported below.

For this generalized volunteer's dilemma with shared rewarding, the average payoffs of the strategies Volunteer and Ignore are given by

$$\begin{aligned} P_V &= \sum_{M=0}^{N-1} \binom{N-1}{M} x^M (1-x)^{N-1-M} \cdot [P_V(M+1) + \tfrac{\delta}{M+1}] \\ &= \sum_{M=k-1}^{N-1} \binom{N-1}{M} x^M (1-x)^{N-1-M} \cdot (1-c) \\ &\quad + [1 - \sum_{M=k-1}^{N-1} \binom{N-1}{M} x^M (1-x)^{N-1-M}] \cdot (1-a-c) \\ &\quad + \sum_{M=0}^{N-1} \binom{N-1}{M} x^M (1-x)^{N-1-M} \cdot \tfrac{\delta}{M+1} \end{aligned} \quad (2a)$$

and

$$\begin{aligned} P_I &= \sum_{M=0}^{N-1} \binom{N-1}{M} x^M (1-x)^{N-1-M} P_I(M) \\ &= \sum_{M=k}^{N-1} \binom{N-1}{M} x^M (1-x)^{N-1-M} \cdot 1 \\ &\quad + [1 - \sum_{M=k}^{N-1} \binom{N-1}{M} x^M (1-x)^{N-1-M}] \cdot (1-a), \end{aligned} \quad (2b)$$

where $x$ is the current fraction of volunteers in the population.

## 2.3. Replicator dynamics

For studying the evolutionary dynamics of $x$, we use the replicator equation (Hofbauer and Sigmund, 1998),

$$\dot{x} = x(1-x)(P_V - P_I),$$

where the dot denotes the derivative with respect to time $t$, $\dot{x} = dx/dt$.

In addition to the boundary equilibria at $x = 0$ and $x = 1$, interior equilibria are found by equating $P_V$ and $P_I$,

$$P_V - P_I = \binom{N-1}{k-1} x^{k-1}(1-x)^{N-k} a - c + \delta \frac{1-(1-x)^N}{Nx} = 0.$$

We introduce the reward ratio

$$p = \frac{\delta}{Nc}$$

($0 < p < 1$) to characterize the magnitude of the total reward $\delta$ relative to the maximal total volunteering cost $Nc$, which yields

$$P_V - P_I = \binom{N-1}{k-1} x^{k-1}(1-x)^{N-k} - c\frac{x - p + p(1-x)^N}{x}.$$

The replicator equation can thus be written as

$$\dot{x} = (1-x)F(x), \tag{3a}$$

with

$$F(x) = \binom{N-1}{k-1} x^k (1-x)^{N-k} a - cf(x) \tag{3b}$$

and

$$f(x) = x - p + p(1-x)^N. \tag{3c}$$

Below, we use Eq. (3) to study equilibria in the fraction of volunteers, in particular as a function of the reward ratio, by theoretical and numerical analyses.

## 3. Results

### 3.1. Defection traps

As a baseline for studying the impacts of shared rewarding, we start with results for the effect of the cost ratio $c/a$ on the social dynamics of the $N$-person volunteer's dilemma without shared rewarding ($p = 0$). Fig. 1a shows that, depending on this cost ratio, beside the two boundary equilibria at $x = 0$ (stable) and $x = 1$ (unstable), the population can have two additional, interior equilibria, one stable and the other unstable [3]. In particular, the unstable interior equilibrium, if it exists, divides the range $[0,1]$ of $x$ into two basins of attraction. As the cost ratio increases, the stable interior equilibrium decreases, while the unstable interior equilibrium increases, until they eventually collide at $c/a = (k-1)/(N-1)$ in a saddle-node bifurcation (Archetti and Scheuring, 2011). For higher cost ratios, there are no interior equilibria, and the only stable equilibrium is at $x = 0$, corresponding to the complete absence of volunteers (Fig. 1a).

These results highlight the existence of defection traps in the generalized volunteer's dilemma: for cost ratios below $(k-1)/(N-1)$, a population initially lacking volunteers ($x = 0$) is trapped in that state, even though a stable equilibrium with a considerable fraction of volunteers exists. The stable interior equilibrium, however, can only be reached when the initial fraction of volunteers exceeds the interior unstable equilibrium.

Against this baseline expectation, we can now study the impacts of shared rewarding, depending on the reward ratio $p$. We can distinguish three qualitatively different cases. First, for very small reward ratios ($p < 1/N$), the configuration of equilibria remains unaffected by shared rewarding (Fig. 1d). A corresponding proof is provided in Appendix A. Second, for slightly larger reward ratios, a saddle-node bifurcation occurs, through which two different configurations of equilibria may arise (Appendix A): for large cost ratios $c/a$ the population possesses only one stable interior equilibrium with a small fraction of volunteers, whereas for small cost ratios it also possesses a second stable interior equilibrium with a large fraction of volunteers (Fig. 1g). In either case, the boundary equilibrium $x = 0$ is destabilized by the shared rewarding (Fig. 1g). Third, for even larger reward ratios, there exists only one stable interior equilibrium, at which the fraction of volunteers decreases with increasing cost ratio (Fig. 1j; Appendix A).

We now consider the effect of the threshold number $k$ of volunteers. Without shared rewarding and for $k = 1$, our model recovers the classical volunteer's dilemma (Diekmann, 1985). In this situation, there is only one interior equilibrium at $x = 1 - (c/a)^{1/(N-1)}$ (Archetti and Scheuring, 2011). For $k > 1$, two interior equilibria exist, one stable and the other unstable, both of which increase with increasing $k$. The fraction of volunteers at the stable equilibrium approaches $x = 1$ as $k$ reaches $N - 1$ (Fig. 1b). As for the impact of shared rewarding, the dependence on $k$ of the long-term dynamics of volunteering in the population is not changed as long as the reward ratio $p$ is small (Fig. 1e). In contrast, for larger reward ratios (Fig. 1h,k), small threshold numbers $k$ result in a single interior equilibrium that is stable, whereas larger threshold numbers result in three interior equilibria, two of which are stable. The critical value of $k$ separating these two cases increases with the reward ratio. Fig. 1b,e,h,k highlight how defection traps can be overcome through sufficient shared rewarding, as long as the threshold number of volunteers is not too high.

Furthermore, we investigate the effect of group size $N$. Without shared rewarding, increasing group size decreases the two interior equilibria (Fig. 1c). Shared rewarding results in a stable interior equilibrium at which the fraction of volunteers approaches $p$ as the group size is increased (Fig. 1f,i,l). When the reward ratio $p$ is small, two interior equilibria exist for small groups, one stable and the other unstable (Fig. 1f); for intermediate groups, an additional stable interior equilibrium emerges (Fig. 1f), and for large groups, this collides with the unstable interior equilibrium, leaving only a single stable interior equilibrium (Fig. 1f). When the reward ratio is increased, the aforementioned three interior equilibria exist only for very small groups (Fig. 1i); otherwise, there is a single stable interior equilibrium (Fig. 1i). And when the reward ratio is increased further, there is a single stable interior equilibrium for all group sizes (Fig. 1l).

It is worth pointing out three surprising impacts of altering group size. First, when there are two stable interior equilibria, the lower one gradually increases with group size (Fig. 1f,i). Second, as group size is increased further, the lower stable interior equilibrium disappears in a saddle-node bifurcation and the population abruptly jumps to what was the higher stable interior equilibrium (Fig. 1f,i) and now becomes the only stable interior equilibrium (Fig. 1l). In both cases, therefore, larger group sizes promote volunteering, contrary to what might naively be expected and what applies in the absence of shared rewarding (Franzen, 1995; Weesie and Franzen, 1998; Archetti, 2009b; Brännström et al., 2011). Third, taken together, the two highlighted effects yield a hysteresis effect that can be used to maximize the fraction of volunteers: for groups of intermediate size, the highest stable level of volunteering is reached by first increasing and then decreasing group size (Fig. 1f,i).

### 3.2. Cusp bifurcation

For understanding the core phenomenon underlying the seemingly intricate dependencies revealed above, it is helpful to integrate into a single diagram the dependences of the stationary fraction $x$ of volunteers on the cost ratio $c/a$ on the one hand and on the reward ratio $p$ on the other. The result is shown in Fig. 2a and can be summarized as follows.

First, for high values of $p$, and irrespective of $c/a$, only a single stable interior equilibrium exits, which gradually increases to $x = 1$ as the reward ratio is increased to $p = 1$.

Second, three interior equilibria coexist only at low values of $c/a$ and $p$, highlighted by the grey region in Fig. 2c. As we have explained above, such combinations of $c/a$ and $p$ imply defection traps.

Third, as either $c/a$ or $p$ are increased from within this region, the unstable interior equilibrium collides with one of the two stable interior equilibria: with the upper one as $c/a$ is increased, and with the lower one as $p$ is increased. In Fig. 2c, the two lines along which these collisions happen, through saddle-node bifurcations, are shown in black.

Fourth, where these two lines connect, indicated by the filled circle in Fig. 2a,c, both aforementioned bifurcations occur together. This means that all three interior equilibria collide at one value of $x$, leaving behind a single stable interior equilibrium at that value of $x$, and thus eliminating the defection trap. This occurs for combinations of $c/a$ and $p$ situated right at the cusp of the defection-trap region shown in Fig. 2c, and amounts to what is therefore known as a cusp bifurcation. Consequently, when the reward ratio $p$ exceeds the reward ratio $p_{\text{cusp}}$ at the cusp, defection traps cannot occur for any value of the cost ratio $c/a$. This means that $p_{\text{cusp}}$ determines a critical level of shared rewarding beyond which defection traps in the generalized volunteer's dilemma are safely overcome.

As long as the group size $N$ exceeds the threshold number $k$, $p_{\text{cusp}}$ remains smaller than 1. Fig. 2b-j show how the defection-trap region expands, and $p_{\text{cusp}}$ increases, with increasing $k$ and decreasing $N$.

### 3.3. Critical reward ratio

While adopting the reward ratio $p_\text{cusp}$ safely eliminates defection traps *for any level* of the cost ratio $c/a$, thus the prescribed reward ratio is overly conservative: this is because except at the cusp itself the critical reward ratio $p_\text{c}$ that eliminates defection traps *for a given level* of the cost ratio $c/a$ is always lower than the reward ratio at the cusp, $p_\text{c} < p_\text{cusp}$.

In Fig. 3a-c, we therefore examine how the critical reward ratio $p_\text{c}$ varies with the cost ratio $c/a$, the threshold number $k$ of volunteers, and the group size $N$. We find that $p_\text{c}$ increases with $k$, $c/a$, and $N$. This is compatible with an understanding that producing a public good is more difficult when it requires a higher threshold number of volunteers or a higher cost ratio (Souza et al., 2009; Boza and Számadó, 2010; Archetti and Scheuring, 2011), or when larger group size inhibits the emergence of volunteering (Franzen, 1995; Weesie and Franzen, 1998; Archetti, 2009b). Moreover, for many combinations of $c/a$, $k$, and $N$, the critical reward ratio $p_\text{c}$ is relatively small compared to the total cost of contributing to the public good, as measured by the cost ratio $c/a$. This opens up opportunities for overcoming defection traps at little extra cost invested into shared rewarding.

To better understand the beneficial effects of shared rewarding, Fig. 3d-f show how the fraction of volunteers varies with $c/a$, $k$, and $N$. We find that the fraction of volunteers decreases with $c/a$ and $N$, but increases with $k$. Moreover, for many combinations of $c/a$, $k$, and $N$, the fraction of volunteers is relatively high. The aforementioned opportunities for overcoming defection traps thus go hand in hand with possibilities of achieving relatively high levels of volunteering, all at little extra cost invested into shared rewarding.

To further explore these options for garnering high returns on small extra investments, we define the relative gain of shared rewarding as the proportion between the achieved fraction of volunteers and the required reward ratio. Fig. 3g-i show how this consideration helps identify windows of opportunity for cost ratios $c/a$ that are smaller than about 0.4 and for threshold ratios $k/N$ that are smaller than about 0.5. Within these windows, shared rewarding is particularly efficient.

## 4. Discussion

In this work, we have incorporated shared rewarding into the generalized $N$-person volunteer's dilemma, and have studied for the first time the resultant evolutionary dynamics of volunteering in infinite well-mixed populations. We have found that small reward ratios, on the order of the inverse of the group size, already enable the persistence of volunteers.

At intermediate reward ratios, the evolutionary dynamics are bistable and exhibit a hysteresis. In this regime, it is possible to move from a state with a low proportion of volunteers to a state with a high proportion of volunteers by first increasing the group size beyond a certain threshold and then decreasing it back to the original level.

When the reward ratio is increased further beyond a critical value, defection traps disappear and the evolutionary dynamics always approach a state with a high proportion of volunteers. For a wide range of parameter combinations, we observe that the critical value of the reward ratio is reached while the amount of rewarding is still relatively small in comparison to the value of the public good that is produced. This highlights shared rewarding as a powerful mechanism for promoting voluntary contributions to the common good of societies.

### 4.1. Comparison with related studies

It is worthwhile to point out that, in the generalized $N$-person volunteer's dilemma, the public good is produced or safeguarded as a step function of individual contributions. This function thus differs from the linear function used in some previous works (Cuesta et al., 2008; Jiménez et al., 2008; Jiménez et al., 2009) on the $N$-person prisoner's dilemma game. In comparison to the case considered here, the $N$-person prisoner's dilemma can be considered as an extreme case, which — due to the assumed linearity — cannot capture the effects of thresholds for the production of the public good. This explains the significantly higher levels of rewards that are needed to establish a significant cooperation level in the $N$-person prisoner's dilemma game (Cuesta et al., 2008). The generalized $N$-person volunteer's dilemma is a good approximation of general public goods games with nonlinear functions, e.g., the sigmoid function (Boza and Számadó, 2010; Frank, 2010; Archetti and Scheuring, 2011), in well-mixed populations. As in those general public goods games, we find that there exist defection traps in the generalized volunteer's dilemma which, as we show, shared rewarding can effectively overcome.

The effects of rewarding on cooperation have already been investigated in several earlier studies (Sigmund et al., 2001; Rand et al., 2009; Hauert, 2010; Hilbe and Sigmund, 2010; Szolnoki and Perc, 2010; Szolnoki and Perc, 2012; Milinski and Rockenbach, 2012). However, these studies mainly examined peer rewarding, with rewarding contributors paying a personal cost to provide contributors with an additional benefit. Rewarding cooperation is expensive in a population of contributors, but is cheap when everyone defects; hence, such rewarding can encourage cooperation. In the case of peer rewarding, players contributing to the common good but refusing to provide rewards achieve higher payoffs than players that reward other players. They thus thwart the attempts to sustain cooperation based on rewards, implying a second-order rewarding dilemma (Hauert, 2010; Hilbe and Sigmund, 2010). In contrast to these earlier studies, the rewarding strategy we have considered here is not costly to players, and correspondingly, players cannot prevent rewards being distributed to volunteers as positive incentives. Hence, the shared rewarding considered here constitutes a form of institutional rewarding (Isakov and Rand, 2012; Cressman et al., 2012; Sasaki et al., 2012), which naturally avoids the emergence of second-order free-riders. Let us emphasize that even if all players need to pay an "entrance fee" (Sasaki et al., 2012) for enabling shared rewarding (so that each player's payoff is reduced by $pc$), the evolutionary outcomes reported here will not be affected.

### 4.2. Limitations and extensions

In this work, we have studied shared rewarding under a variety of simplifying assumptions that could be relaxed or extended. First, following the traditional setting of evolutionary game theory, we have considered infinite well-mixed populations. Hence, a natural extension of our analyses is to consider the effects of finite population size (Pacheco et al., 2009; Souza et al., 2009). Also, the proposed model does not include spatial structure (Sigmund et al., 2001), voluntary participation (Hauert et al., 2002), reputation (Hauert, 2010), or exploratory behavior (Traulsen et al., 2009). In such more complex scenarios, the positive effect of shared rewarding, leading to the avoidance of defections traps, may be further amplified.

Second, we have assumed that volunteers can always receive a certain amount of reward from an outside source, no matter whether the public good at stake is produced or not. However, in some situations volunteers receive such a reward only if they produce the public good. Hence, a promising extension of this work is to consider alternative funding structures for shared rewarding.

Third, we have assumed that the total reward offered to the group of volunteers is equally shared within that group. As an alternative, it would be interesting to consider such rewards being distributed in proportion to the magnitude of contribution made by each volunteer. Naturally, this requires employing an extended theoretical framework in which contributions made to a public good are described as continuous variables (Doebeli and Hauert, 2005), instead of treating them are merely binary choices.

Fourth, here we have examined the repercussions of shared rewarding assuming a threshold rule for the considered public good, which is either produced or not, depending on whether the number of volunteers exceeds a threshold. It will be interesting to relax this simplifying assumption, for example, by replacing the considered step function with a sigmoidal function.

Fifth, in view of recent advances concerning punishments, it remains to be seen how the findings we have reported are affected by the joint impacts of rewarding and punishing (Hilbe and Sigmund, 2010; Chen et al., 2013). It will also be worthwhile to explore the implications of nonlinear cost-benefit functions and spatially structured populations (Szolnoki and Perc, 2010; Szolnoki and Perc, 2012).

### Acknowledgments

UD gratefully acknowledges financial support by the Austrian Science Fund, through a grant for the research project *The Adaptive Evolution of Mutualistic Interactions* as part of the multinational collaborative research project *Mutualisms, Contracts, Space, and Dispersal (BIO-*

*CONTRACT*) selected by the European Science Foundation as part of the European Collaborative Research (EUROCORES) Programme *The Evolution of Cooperation and Trading (TECT)*. Additional funding to UD was provided by the European Commission, the European Science Foundation, the Austrian Ministry of Science and Research, and the Vienna Science and Technology Fund.## Appendix A: Analysis of equilibria

The replicator dynamics in Eq. (3) have two boundary equilibria: $x = 0$ and $x = 1$. Since the first term of $F(x)$ is positive for all $x \in (0, 1)$, interior equilibria can exist only at frequencies $x$ at which $f(x) \geq 0$ (The interior equilibria of these dynamics are determined by the interior roots of $F(x)$, as defined in Eq. (3b)). If $Np \leq 1$, $f(x) > 0$ for all $x \in (0,1)$; otherwise, $f(x)$ has a single interior root $\bar{x} \in (0,1)$, with $f(x) > 0$ for $x \in (\bar{x}, 1)$ (Appendix B).

The analysis of the interior roots of $F(x)$ can be reduced to studying the slope and curvature of the function $G(x)$, defined as

$$G(x) = \binom{N-1}{k-1} x^k (1-x)^{N-k} / f(x),$$

since $F(x) = 0$ is equivalent to $G(x) = c/a$. We thus compute

$$G'(x) = \binom{N-1}{k-1} x^{k-1}(1-x)^{N-k-1} D(x)/f^2(x),$$

where

$$D(x) = x[x - 1 + N(p - x)] + k[x - p + p(1-x)^N].$$

This shows that the sign of $G'(x)$ is that of $D(x)$, which implies that the roots of $G'(x)$ and of $D(x)$ are identical. For later reference, we note that the first derivative of $D(x)$ is given by

$$D'(x) = k - 1 + N(p - 2x) - kNp(1-x)^{N-1} + 2x,$$

and the second derivative of $D(x)$ is given by

$$D''(x) = (N-1)[kNp(1-x)^{N-2} - 2].$$

Based on these preparations, we now study the interior equilibria of Eq. (3) in two situations.

### (a) Insufficient rewarding, $p < 1/N$

We use that $f(x) > 0$ for all $x \in (0, 1)$ (Appendix B): thus, any interior equilibria of Eq. (3) must be located in $(0,1)$.

When $kNp \leq 2$, we need to distinguish two different cases. First, when $k > 1$, $D'(0) = (k-1)(1-Np) > 0$ and $D(0) = 0$. Hence, $D(x)$ is positive near $x = 0$. Since $D(1) = (1-p)(k-N) < 0$, and $D''(x) < 0$, $D(x)$ has a unique interior root in $(0,1)$.

Page 12 of 22

Second, when $k = 1$, $D'(0) = 0$. Since $D''(x) < 0$ and $D(1) < 0$, $D(x) < 0$ for $0 < x < 1$; thus, $D(x)$ has no interior roots in $(0,1)$.

When $kNp > 2$, $D''(x) = 0$ for one $x \in (0,1)$, so $D(x)$ has a unique inflection point at $x = \tilde{x} = 1 - (2/(kNp))^{\frac{1}{N-2}}$. Accordingly, $D''(x) > 0$ for $0 < x < \tilde{x}$, and $D''(x) < 0$ for $\tilde{x} < x < 1$. $D(x)$ is positive near $x = 0$ (since $D'(0) = (k-1)(1 - Np) > 0$ and $D(0) = 0$), and $D(1) < 0$, hence $D(x)$ has a unique interior root in $(0,1)$.

We can thus conclude that for $k > 1$ $G'(x)$ and $D(x)$ have a unique interior root $\hat{x}$, and $G(\hat{x})$ is the unique interior maximum of $G(x)$. Accordingly, solving the equation $G(x) = c/a$ leads to the following conclusions:

(1) When $G(\hat{x}) > c/a$, Eq. (3) has two interior equilibria, denoted by $x_1^*$ and $x_2^*$ with $x_1^* < x_2^*$. Since $\hat{x}$ is the unique interior root of $D(x)$, and since $D(x)$ is positive near $x = 0$, $D(x)$ is positive for $0 < x < \hat{x}$ and negative for $\hat{x} < x < 1$. This implies that $x_1^*$ is an unstable equilibrium and $x_2^*$ is a stable equilibrium.

(2) When $G(\hat{x}) = c/a$, Eq. (3) has only one interior equilibrium $\hat{x}$, which is a tangent point, and thus unstable.

(3) When $G(\hat{x}) < c/a$, Eq. (3) has no interior equilibria.

When $k = 1$, $D(x)$ has no interior roots in $(0,1)$ and $D(x) < 0$, so $G(x)$ is monotonically decreasing. Therefore, Eq. (3) has only a single stable interior equilibrium in $(0,1)$.

We can thus conclude that for $Np < 1$ the interior equilibria of Eq. (3) have the same structure (i.e., topology and stability) as those of the generalized volunteer's dilemma without shared rewarding (Archetti and Scheuring, 2011).

### (b) Sufficient rewarding, $p > 1/N$

We use that $f(x) \leq 0$ for all $x \in (0, \bar{x}]$ (Appendix B): thus, $F(x)$ is positive for $0 < x \leq \bar{x}$, and any interior equilibria of Eq. (3) must be located in $(\bar{x}, 1)$.

When $kNp \leq 2$, $D''(x) < 0$ for $\bar{x} < x < 1$. Since $D(\bar{x}) < 0$ for $Np > 1$ (Appendix C) and $D(1) = (1-p)(k-N) < 0$, $D(x)$ has at most two roots in $(\bar{x}, 1)$.

When $kNp > 2$, we need to distinguish two different cases. First, when $\bar{x} \geq \tilde{x}$, $D''(x) < 0$ for $\bar{x} < x < 1$. Thus, $D(x)$ again has at most two roots in $(\bar{x}, 1)$. Second, when $\bar{x} < \tilde{x}$ and $D(\tilde{x}) \geq 0$, $D(x)$ has a single root in $(\bar{x}, \tilde{x}]$ (since $D''(x) > 0$ and $D(\bar{x}) < 0$), and a single root in $(\tilde{x}, 1)$ (since $D''(x) < 0$ and $D(1) < 0$); thus, $D(x)$ has two roots in $(\bar{x}, 1)$. When $\bar{x} < \tilde{x}$ and $D(\tilde{x}) < 0$, $D(x)$ has no interior roots in $(\bar{x}, \tilde{x})$, and at most two roots in $(\tilde{x}, 1)$ (Since $D''(x) < 0$ and $D(1) < 0$); thus, $D(x)$ has at most two roots in $(\bar{x}, 1)$.

We can thus conclude that for $Np > 1$ $D(x)$ has at most two roots in $(\bar{x}, 1)$, denoted by $\check{x}$ and $\hat{x}$ with $\check{x} < \hat{x}$. We then have $D(x) < 0$ for $x \in (\bar{x}, \check{x})$, $D(x) > 0$ for $x \in (\check{x}, \hat{x})$, and $D(x) < 0$ for $x \in (\hat{x}, 1)$. Thus, $G(\check{x})$ is a minimum of $G(x)$ and $G(\hat{x})$ is a maximum of $G(x)$. In addition, $f(\check{x}) > 0$ since $\check{x} > \bar{x}$, and $f(x)$ approaches zero from above for $x \to \bar{x}$ (Appendix B). As a result, we have $\lim_{x \to \bar{x}^+} G(x) > G(\hat{x}) > G(\check{x}) > G(1) = 0$. Accordingly, solving the equation $G(x) = c/a$ leads to the following conclusions (Fig. 1g):

(1) When $G(\check{x}) > c/a$ or $G(\hat{x}) < c/a$, Eq. (3) has only one interior equilibrium, denoted by $x_1^*$. This equilibrium is stable, since $D(x_1^*) < 0$.

(2) When $G(\hat{x}) = c/a$ or $G(\check{x}) = c/a$, Eq. (3) has two interior equilibria, denoted by $x_1^*$ and $x_2^*$ with $x_1^* < x_2^*$. Specifically, for $G(\hat{x}) = c/a$, $x_1^*$ is stable, since $D(x_1^*) < 0$, and $x_2^* = \hat{x}$ is a tangent point (Pacheco et al., 2009; Souza et al., 2009). In contrast, for $G(\check{x}) = c/a$, $x_2^*$ is stable, since $D(x_2^*) < 0$, and $x_1^* = \check{x}$ is a tangent point.

(3) When $G(\check{x}) < c/a < G(\hat{x})$, Eq. (3) has three interior equilibria, denoted by $x_1^*$, $x_2^*$, and $x_3^*$ with $x_1^* < x_2^* < x_3^*$. Of these, $x_1^*$ and $x_3^*$ are stable, since $D(x_1^*) < 0$ and $D(x_3^*) < 0$, whereas $x_2^*$ is unstable, since $D(x_2^*) > 0$.

When $D(x)$ has only one root or no roots in $(\bar{x}, 1)$, $G(x)$ is monotonically decreasing, since $D(x) \leq 0$. Therefore, Eq. (3) has only a single stable interior equilibrium in $(\bar{x}, 1)$ (Fig. 1j).

## Appendix B: Analysis of the function $f(x)$

From the expression of $f(x)$ in Eq. (3c), we see that $f(0) = 0$ and $f(1) = 1 - p > 0$. Moreover, $f'(x) = 1 - Np(1-x)^{N-1}$ and $f''(x) = (N-1)Np(1-x)^{N-2}$. Thus, $f''(x) > 0$ for all $x \in (0,1)$.

On this basis, we can draw the following conclusions for $Np \leq 1$ and $Np > 1$. First, for $Np \leq 1$, $f'(x) > 0$, so $f(x)$ is positive for all $x \in (0,1)$. Second, for $Np > 1$, $f'(0) = 1 - Np < 0$. Hence, $f(x)$ is negative near $x = 0$. Together with $f(1) > 0$ and $f''(x) > 0$ for all $x \in (0,1)$, this means that $f(x)$ has a single interior root $\bar{x}$ in $(0,1)$. In addition, we have $f(x) > 0$ for all $x \in (\bar{x}, 1)$. We also know that $\bar{x} < p$, since $f(p) = p(1-p)^N > 0$.

## Appendix C: Analysis of the sign of $D(\bar{x})$

To determine the sign of $D(\bar{x}) = \bar{x}[Np - 1 - (N-1)\bar{x}] + kf(\bar{x})$ with $f(\bar{x}) = 0$, we need to know the relationship between $\bar{x}$ and $\check{x} = (Np-1)/(N-1)$. According to Appendix B, $\check{x} < \bar{x}$ is equivalent to $f(\check{x}) < f(\bar{x}) = 0$, so $D(\bar{x})$ has the same sign as
$$f(\check{x}) = \frac{p-1}{N-1} + p(1-p)^N \left(\frac{N}{N-1}\right)^N.$$

To evaluate the sign of $f(\check{x})$, we define the continuous function $h(p) = f(\check{x})$, and determine $h'(p)$ and $h''(p)$ as
$$h'(p) = \frac{1}{N-1} - \left(\frac{N}{N-1}\right)^N (1-p)^{N-1}(Np + p - 1)$$
and

$$h''(p) = N\left(\frac{N}{N-1}\right)^N (1-p)^{N-2}(Np + p - 2).$$

We thus see that $h(p)$ has roots at $p = \hat{p} = \frac{1}{N}$ and $p = 1$, and has the unique maximum at $\hat{p}$ for $\frac{1}{N} \leq p < 1$ (since $h'(\frac{1}{N}) = 0$ and $h''(\frac{1}{N}) < 0$). Hence, $h(p) < 0$ for $p \in (\frac{1}{N}, 1)$, and thus, $D(\bar{x}) < 0$ for $p > \frac{1}{N}$.

# Figure captions

**Figure 1.** Stationary fraction of volunteers. Shown are the stable equilibria (thick lines) and unstable equilibria (thin lines) as a function of the cost ratio $c/a$ (left column), the threshold number $k$ of volunteers (centre column), and the group size $N$ (right column) for different values of the reward ratio $p$ (rows). Between the equilibria, the fraction of volunteers either increases (grey) or decreases (white) as a result of the evolutionary dynamics (this is also indicated by arrows). Even without rewarding, a stable equilibrium exists with volunteers constituting is a large fraction of the population (top row). However, depending on initial conditions, the population may also approach the boundary equilibrium, where the fraction of volunteers equals zero. As the reward ratio is increased (second row), a second interior equilibrium emerges, as the state of full defection becomes unstable (f). Increasing the reward ratio further reduces the range of parameter combinations for which population converges to a low proportion of volunteers, so that ultimately the only stable equilibrium is a state with a high proportion of volunteers (fourth row). Parameters: $N = 20$ and $k = 10$ in (a), (d), (g), and (j); $N = 20$ and $c/a = 0.12$ in (b), (e), (h), and (k); $k = 10$ and $c/a = 0.12$ in (c), (f), (i), and (l).

**Figure 2.** Avoidance of defection traps due to a cusp bifurcation. Top: The interior equilibria of the system can be visualized as forming a single continuous surface in the three-dimensional space spanned by the cost ratio $c/a$, the reward ratio $p$, and the equilibrium fraction of volunteers. For most parameter combinations $(c/a, p)$, there is a unique stable equilibrium. However, when cost ratio and reward ratio are both small (front corner), the surface folds back onto itself and three interior equilibria are formed accordingly, of which the top and bottom ones are stable which the intermediate one is unstable, resulting in bistability. The fold lines correspond to the parameter combinations at which saddle-node bifurcations occur: at these, stable and unstable equilibria collide and annihilate each other. The lines of these saddle-node bifurcations thus mark the boundaries of the bistable region in parameter space. The two saddle-node bifurcation lines themselves collide in a cusp bifurcation (black circle). This cusp thus corresponds to the highest values of the cost and reward ratios at which the defection trap, corresponding to the lower stable equilibrium, exists. Also shown is the lower boundary equilibrium (bottom surface), whereas the upper boundary equilibrium has been omitted for clarity. Bottom: Dependence of cusp location (black circle) and bistable region (grey area) on cost and reward ratios, as a function of group size (rows) and threshold number of volunteers (columns). Shown are two-parameter bifurcation diagrams corresponding to a top view of panel (a) (the saddle-node bifurcations delineating the bistable region are shown as black lines in these figures). As the threshold number of volunteers is increased and/or the group size is decreased, the cusp bifurcation (black circle) moves to higher values of the cost and reward ratio, and the bistable region (grey area) increases in size.

**Figure 3.** Differential performance of shared rewarding, depending on parameter values. Color-coded is the critical reward ratio (top row), the equilibrium fraction of volunteers (middle row), and the relative gain of shared rewarding, computed as the equilibrium fraction of volunteers divided by the critical reward ratio (bottom row), for different values of the group size (columns). Grey areas indicate parameter combinations for which the critical reward tio does not exist. Throughout most of the parameter space, the critical reward ratio is small compared to the total cost of contributing to the public good, and the resultant equilibrium fraction of volunteers is high. Shared rewarding is most efficient in the bottom row's green regions.

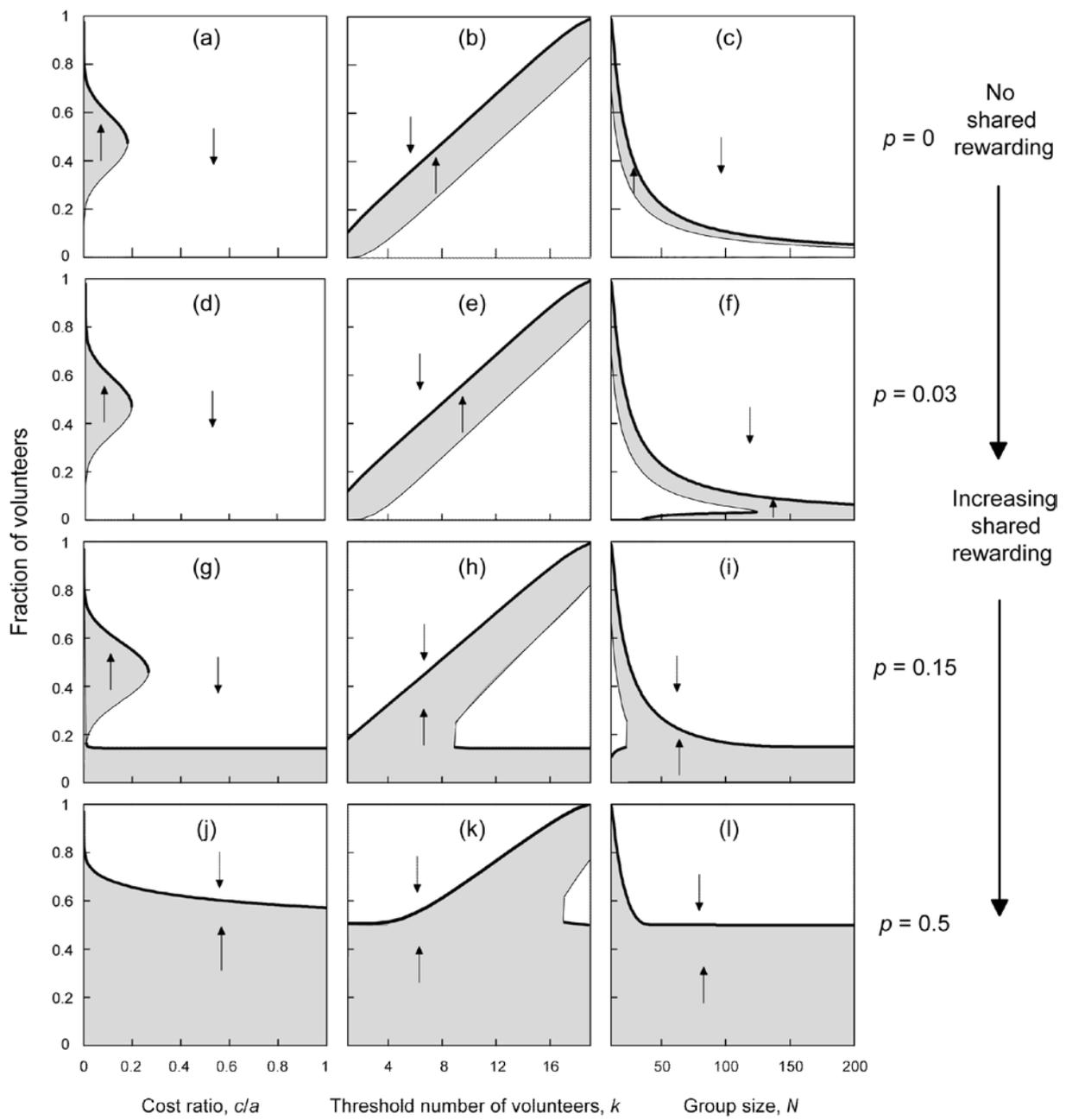

**Figure 1:**

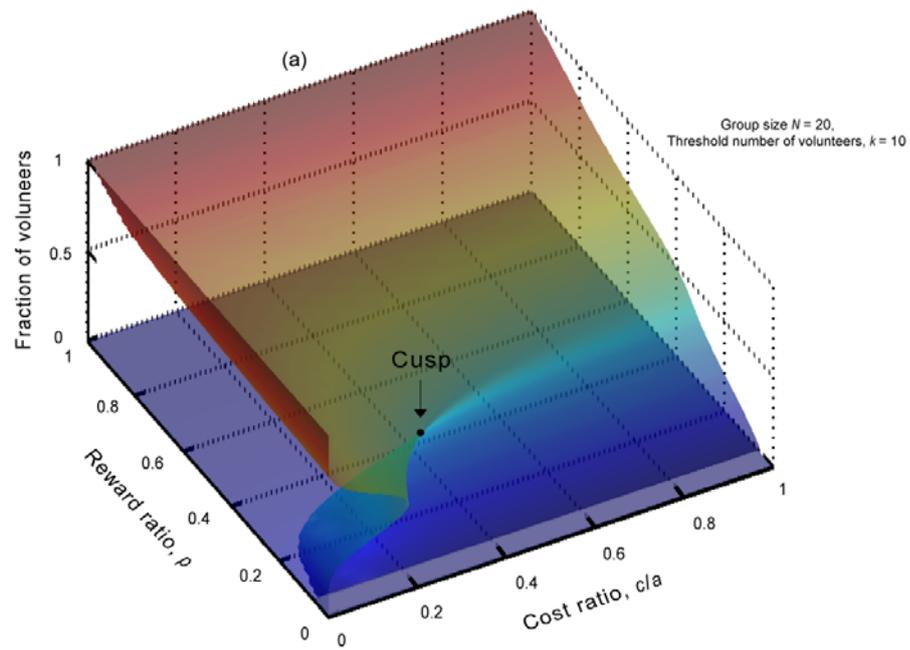
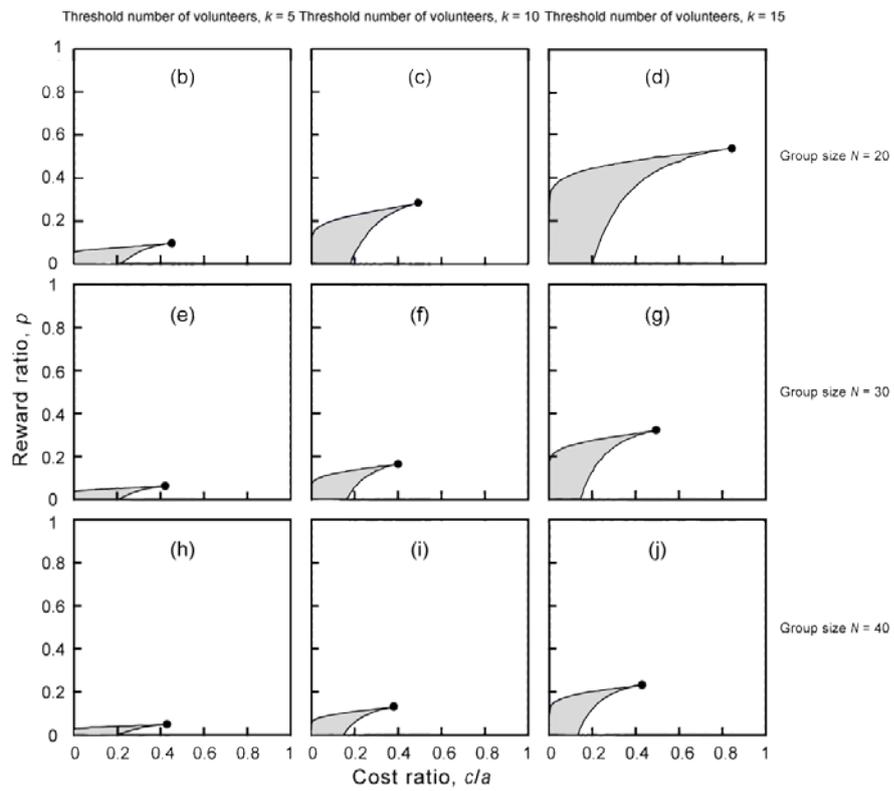

**Figure 2:**

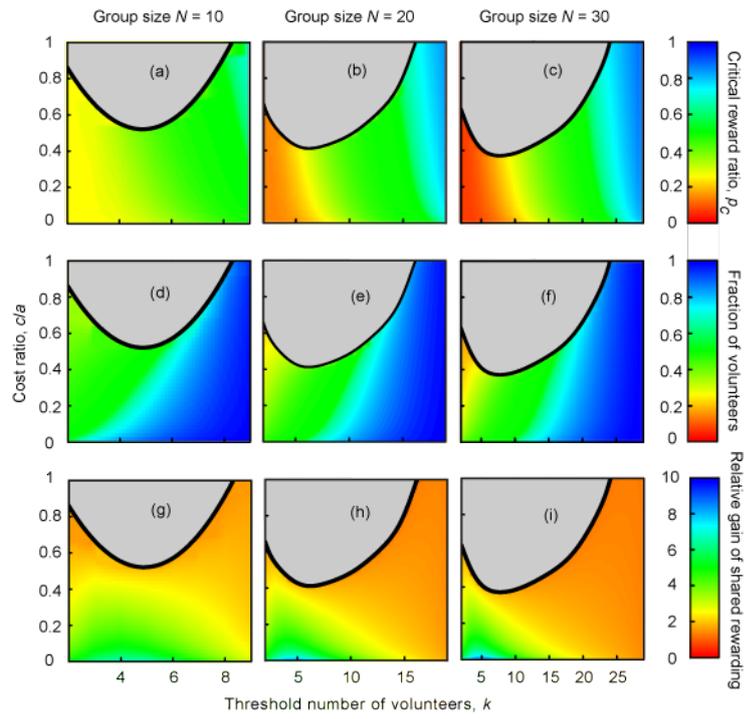

**Figure 3:**